# Feedback-optimized Extraordinary Optical Transmission of Continuous-variable Entangled States

Dong Wang[1,2], Chuanqing Xia[1], Qianjin Wang[1], Yang Wu[1], Fang Liu[1], Yong Zhang[1,*], and Min Xiao[1,3,†]

*1 National Laboratory of Solid State Microstructures, College of Engineering and Applied Sciences, School of Physics, Nanjing University, Nanjing 210093, China*
*2 School of Mathematics and Physics, Anhui University of Technology, Maanshan, 243032, China*
*3 Department of Physics, University of Arkansas, Fayetteville, Arkansas 72701, USA*

**Abstract**
We report on the feedback-optimized extraordinary optical transmission of continuous-variable entangled states through a hexagonal metal-hole array. The continuous-variable entanglements from a nondegenerate optical parametric amplifier are first demonstrated to survive after a photon-plasmon-photon conversion process. By controlling the reflected light from the metal-hole array, a significant enhancement of quantum correlations has then been experimentally achieved comparing to the case of without such coherent feedback control. This result presents a useful technique to efficiently recover the substantial reflective losses in the plasmonic circuits for quantum information processing.

[*]zhangyong@nju.edu.cn
[†]mxiao@uark.edu

In recent years, quantum states of light and surface plasmons are gradually combined to form a new research field of quantum plasmonics [1-7]. This merge has been mostly motivated by building more compact integrated photonic circuits for quantum information processing. The light-excited plasmon modes can be confined into a nanoscale mode volume [5-7], which provides a potential route to overcome the incompatibility between photons and nanoscale circuits because photons easily break free when confined into a sub-wavelength size volume. Also, the localized fields in metal structures can greatly enhance their interactions with materials for further detection of the quantum states [6, 7].

The first important issue to realize such quantum plasmonic circuits is whether the quantum correlations embedded in lights will still be preserved after a photon-plasmon-photon conversion process. Despite the decoherence caused by collisions among millions of electrons that constitute the plasmons, Alterwischer *et al.* had demonstrated a plasmon-assisted transmission of polarization-entangled photons in a metal hole array in 2002 [8]. Later, more experiments using probabilistically-prepared quantum states have shown that energy-time entanglement [9], high-dimensional orbital angular-momentum entanglement [10] and indistinguishability [11] can all be basically preserved after propagating through metal hole arrays or metal nanowires. However, quantum plasmons excited by deterministically-prepared quantum sources, which are usually continuous variable (CV) quantum states described in an infinite-dimension Hilbert space, are less investigated. To the best of our knowledge, only the preservations of quadrature squeezed vacuum states [12] and intensity-difference squeezed states [13] were experimentally demonstrated. So far, the properties of CV-entangled plasmons remain to be unexplored. The unique quantum performance of CV-entangled states makes them extremely useful in quantum information processing, such as quantum key distribution, dense coding, teleportation, entanglement swapping, and computation [14]. Besides, CV entanglement is associated with the quadrature noise correlations of two beams, which can excite quantum plasmons quite differently comparing to the previously reported discrete-variable entangled plasmons [8]. Therefore, it is necessary to study the coupling between such CV-entangled states and plasmons for future integrated circuits applications. In this Letter, we first experimentally demonstrate that CV entanglements from a nondegenerate optical parametric amplifier (NOPA) can be well preserved after extraordinary optical transmission (EOT) through a hexagonal metal-hole array. Periodic array of holes in a metal film is used to covert light into plasmons by providing the necessary momentum conservation for the coupling process, which can give rise to the EOT phenomenon [15,16].

The second important task is to overcome the non-negligible losses in the plasmonic circuits for maximally preserving quantum information during the photon-plasmon conversion because quantum correlations, for example, the CV entanglements, are very sensitive to losses. One obvious way is to design the metal structure to transmit as much quantum light as possible, which, however, has its limit because in most cases a metal structure for realizing certain functionality is impossible to be designed also with a high transmissivity. For example, a transmissivity of 80% in a two-dimensional EOT metal structure is already hard to realize in experiment. But a 20% loss is still detrimental for quantum information processing. By carefully analyzing the loss channels when coupling light into a metal structure, one realizes that high reflectivity at the metal surface contributes to a considerable proportion of the total "loss". Here, we propose and experimentally demonstrate a coherent-feedback-control method [17,18] to efficiently reuse the reflected light for maximally preserving the quantum correlations in coupling CV-entangled states of light and plasmons in metal structures. Coherent feedback control is a non-measurement-based control method without any back-action noises induced by the measurements, which makes it very suitable to optimize the quantum correlations in CV quantum information processing.

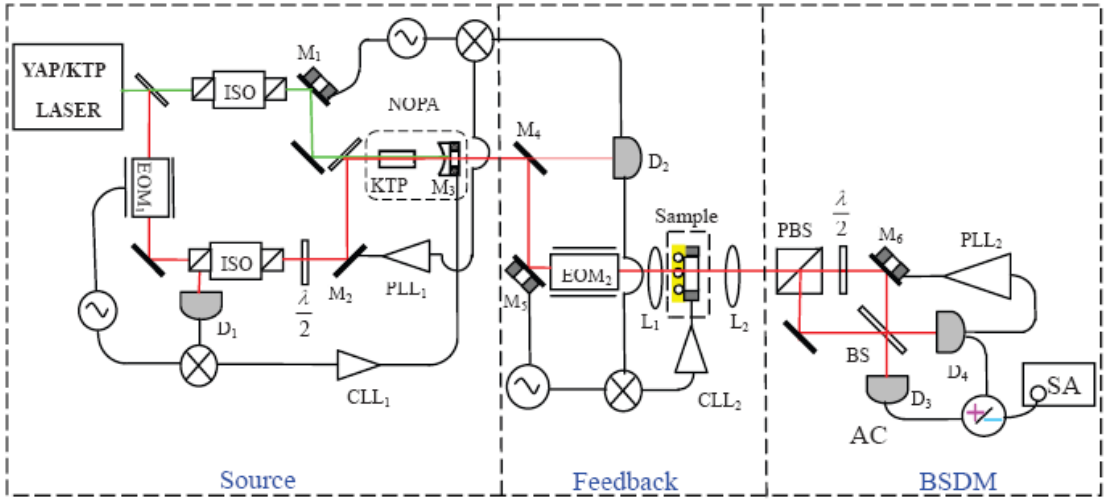

FIG. 1 Experimental schematic diagrams. The CV-entangled states are generated from a NOPA cavity pumped by an intracavity-frequency-doubled 540 nm/1080 nm YAP-KTP laser. The feedback cavity, consisting of the mirror $M_3$ and the sample, is locked on resonance. The entanglement is measured by the BSDM method. $CLL_1$ and $CLL_2$ are the cavity-lock loops for the NOPA and the feedback cavity, respectively. $PLL_1$ and $PLL_2$ are the phase-difference-lock loops for the NOPA and BSDM, respectively. ISO: optical isolator. EOM: electro-optical phase modulator. BS: 50:50 beam splitter. PBS: polarization beam splitter. SA: Spectrum analyzer. D: detector. M: mirror. $\lambda/2$: $\lambda/2$ wave plate

In the experiment, we use a NOPA operating at the deamplification mode to generate the CV-entangled states. As shown in Fig. 1 (Source part), the intracavity-frequency-doubled YAP-KTP laser outputs 220 mW of green light at 540 nm and 12 mW of near-infrared light at 1080 nm, which serve as the pump and seed beams for the NOPA, respectively. The semi-monolithic NOPA cavity consists of an $\alpha$-cut KTP crystal of 10 mm length and an output mirror ($M_3$ in Fig. 1) with 4% transmissivity. By adjusting the temperature of the KTP crystal, 1080 nm and 540 nm lights can satisfy the type-II noncritical phase-matching condition. The phase difference between these two light beams is locked to be π, i.e. the NOPA operates under the parametric-deamplification condition. Two collinear 1080 nm beams with their polarizations perpendicular to each other are generated from the NOPA, which are CV entangled with the anticorrelation of amplitude quadratures and the correlation of phase quadratures [19].

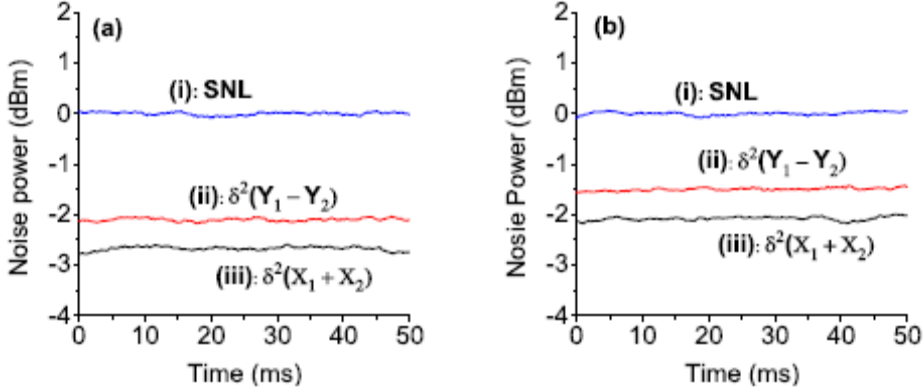

FIG. 2 Correlation noise power spectra of the entangled states at 2 MHz: (a) without the sample and (b) with the sample. Curve (i) is the SNL, (ii) is the phase-difference noise spectrum and (iii) is the amplitude-sum noise spectrum.

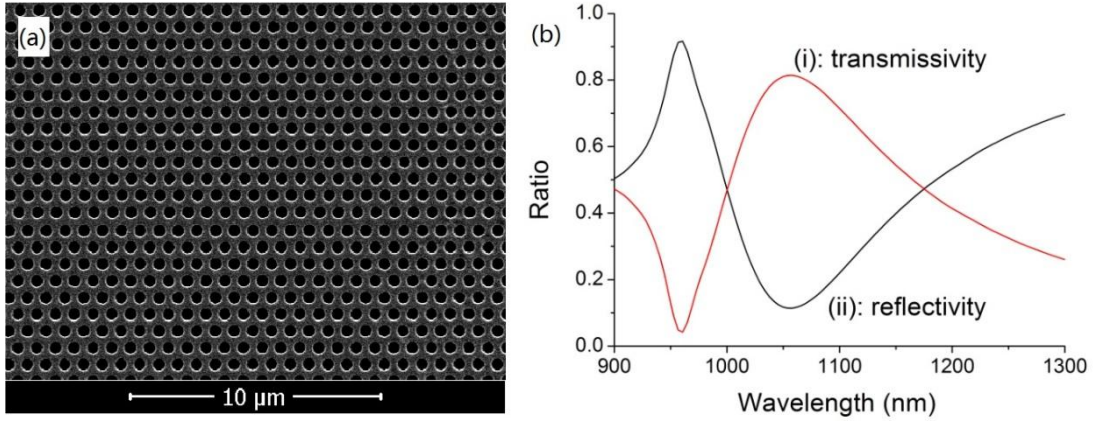

FIG. 3 (a) SEM image of the hexagonal metal-hole array. (b) Simulated reflectivity and transmissivity at different wavelengths.

Through a simple Bell state direct measurement (BSDM) [20], the entanglement of the generated 1080 nm beams was first measured without inserting the metal hole array sample. As shown in Fig. 1, the output lights from the NOPA were split by a polarization beam splitter (PBS). A mirror $M_6$ mounted on a piezoelectric transducer (PZT) was used to lock the phase difference between the two beams to be $\pi/2$ for BSDM. One beam passed through a $\lambda/2$ waveplate and interfered with the other one on a 50:50 beam splitter, which were detected by detectors $D_3$ and $D_4$ (Fig. 1) with AC and DC outputs. The DC signals were used to monitor the light powers and to control the mirror $M_6$ for BSDM. The AC outputs were used to measure the noises of the optical fields. By performing addition or subtraction between the AC signals, we can get the amplitude-sum noise or phase-difference noise of the entangled beams, respectively, through a signal spectrum analyzer (Agilent N9000A). To measure the corresponding shot-noise limit (SNL), we blocked the green light before the NOPA and adjusted the near-infrared powers at detectors $D_3$ and $D_4$ to be the same as the entangled beams. Figure 2(a) shows that the amplitude-sum noise and phase-difference noise of the quantum source were squeezed about 2.7 dB and 2.1 dB, respectively, which satisfy the Duan's inseparability criterion of entanglement [21]

$$\left\langle \delta^2(\frac{X_1+X_2}{\sqrt{2}}) + \delta^2(\frac{Y_1-Y_2}{\sqrt{2}}) \right\rangle = 1.30 < 2. \tag{1}$$

Here, $X_1$ and $X_2$ are the amplitude quadratures, and $Y_1$ and $Y_2$ are the phase quadratures, of the two beams, respectively. The squeezing degree of the phase-difference noise was smaller than that of the amplitude-sum noise because of the laser's extra noise [22].

The metal-hole array for performing the quantum EOT experiments was designed based on two considerations: (1) to maximize the light transmission at 1080 nm; and (2) to realize polarization-independent transmission because the entangled beams in our experiment have two polarizations. A hexagonal hole array can transmit more light than other structures with lower symmetries [23] and its transmission doesn't depend on the polarization of incident light [24]. Using a focused ion beam system (FEI Helios 600i), the hexagonal metal-hole array (Fig. 3(a)) was fabricated in a 58 nm-thick Au film sputtered on a glass substrate. Each circular hole has a diameter of 460 nm and the period of the hexagonal array is 758 nm. The holes only occupy 33.4% of the area in our sample. Thanks to the plasmons, the transmissivity of our sample has been measured to be 75% for both of the polarized 1080 nm beams, which is close to the theoretical value of 80% considering the EOT effect (Fig. 3(b)). Obviously, there is a 25% total loss from the sample. By placing a λ/4 waveplate and a PBS before the sample, the reflective loss was measured to be 14%. The rest 11% can be attributed to scattering and absorption inside the sample. Here, more than half of the total loss comes from the reflection on the metal surface! We will demonstrate that by employing a coherent feedback control, the reflected light can actually be efficiently reutilized to enhance the transmitted quantum correlations.

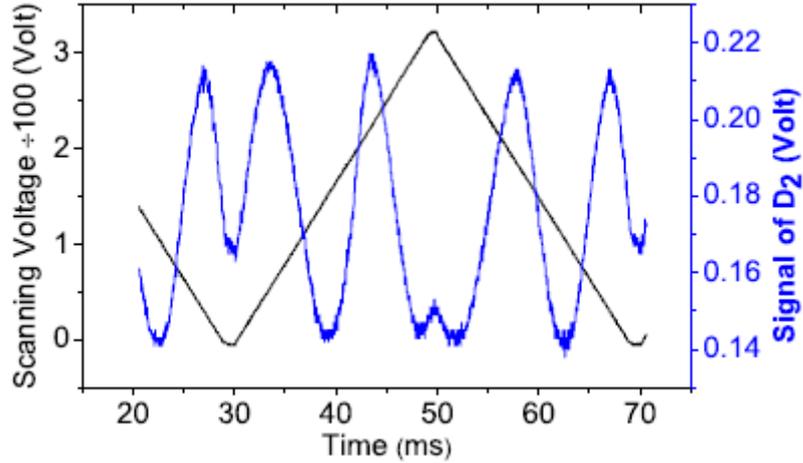

FIG. 4 Dynamical performance of the feedback cavity. The triangle curve in black denotes the scanning voltage on the PZT to drive the sample. The blue curve denotes the power inside the weak feedback cavity.

In the experiment, the metal-hole array was placed at the focal plane of lens $L_1$. The sample surface and the output mirror $M_3$ of the NOPA formed a weak feedback cavity. To make the two orthogonally-polarized 1080 nm beams simultaneously resonate in the feedback cavity, an electro-optical phase modulator ($EOM_2$ in Fig. 1) was used to compensate for the phase difference between them caused by mirrors $M_4$ and $M_5$. The work voltage of $EOM_2$ was set when the output intensity curves (detected by $D_3$ and $D_4$ in Fig. 1) overlapped with the PZT-mounted mirror $M_5$ scanned. The coherent feedback control was realized by tuning the PZT that drives the metal-hole sample. When the feedback cavity was on resonance, the reflected light from the metal sample could hardly enter and affect the NOPA. The output from the NOPA was almost the same as that without the feedback control. In this case, the reflected light was reutilized to enhance the quantum preservation after EOT through the metal-hole sample. If the feedback cavity was not on resonance, more light was reflected back into the NOPA cavity. Then, the quantum output became worsen because the phase of the reflected light did not match the requirement for a deamplified NOPA. However, in an amplified configuration, coherent feedback control can enhance the corresponding quantum properties through the interaction between the reflected light and the nonlinear crystal as reported in Refs. [17,18]. In our experiment, a resonant feedback cavity was the only choice, which was subtly realized by locking the intensity measured at $D_2$ at the lowest value (Fig. 4). Also, a standard Pound-Drever-Hall method [25] was used to keep the NOPA cavity resonant when controlling the feedback. The entanglement of lights passing through the metal-hole array has also been measured by BSDM. The corresponding SNL was determined by blocking the pump light and locking the feedback cavity. As shown in Fig. 2(b), the amplitude-sum and phase-difference noises were measured to be 2.2 dB and 1.6 dB below the SNL, respectively. The sum of the corresponding correlation variances was 1.21 (Eq. (1)), which satisfies the Duan's entanglement criterion [21]. Comparing to the output from the NOPA, the CV entanglement after the feedback-controlled EOT was mostly preserved. The decrease of CV entanglement mainly resulted from the scattering and absorption in the sample.

Next, we analyze the contributions of the EOT effect and the coherent feedback control on the entanglement preservation. EOT is usually defined when the transmission of the array normalized to the area occupied by the holes is larger than 1 [15,16]. Here, we use a similar model for comparison. We assume that without the EOT effect the light could directly pass through the empty holes, which would result in a transmissivity of 33.4% in our sample. From a beam-splitter model [2,13,26], the squeezing of the amplitude-sum noise and the phase-difference noise of the transmitted lights can be calculated by

$$V' = -10\lg(1-T\times 10^{-V/10}) \qquad (2)$$

to be only 0.7 dB and 0.6 dB, respectively, which are far below the experimentally measured results. Here, $V$ is the squeezing of the light noise before the transmission and $T$ is the transmissivity of the metal hole sample. Due to the EOT effect, the transmissivity of our sample was enhanced to be 75%. The degrees of squeezing for the amplitude-sum noise and the phase-difference noise would be 1.9 dB and 1.5 dB, respectively. Clearly, the EOT effect contributed 1.2 dB amplitude-sum squeezing and 0.9 dB phase-difference squeezing. The additionally enhanced amounts of squeezing (0.3 dB for amplitude-sum and 0.1 dB for phase-difference) were clearly resulted from the coherent feedback control.

The classical and quantum characteristics of the plasmon-assisted transmission under coherent feedback control can be calculated by using Langevin equations as reported in Ref. [17]. Here, the cavity detuning, i.e., the phase shift of the 1080 nm light during a round trip in the feedback cavity, needs to be considered. It is easy to see from Fig. 5(a) that the intensity at $D_2$ is lowest when the detuning from resonance is 0, which confirms that the feedback cavity was indeed locked on resonance during the experiment. Using the Langevin equations, we have also calculated the amplitude-sum squeezing of the transmitted light from EOT samples with different reflectivities as shown in Fig. 5(b). The 0.3 dB squeezing enhancement predicted at 14% reflectivity is well consistent with the experimental result. Considering the 70% measurement efficiency for the BSDM, the actual enhancement of the quantum correlation due to feedback control is 15% (0.6 dB). Moreover, the contribution of coherent feedback control will be more significant with an EOT sample having a higher reflectivity as shown in Fig. 5(b).

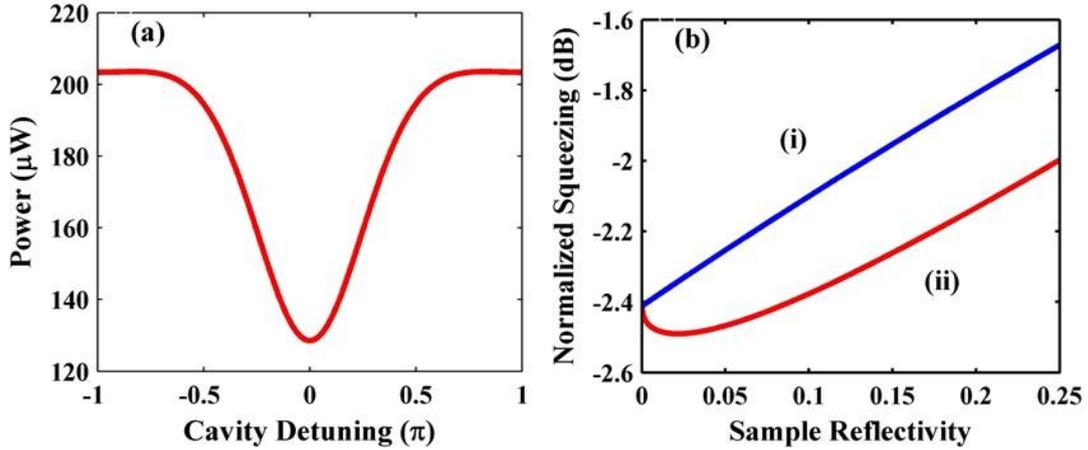

FIG. 5 Simulations using Langevin equations reported in Ref. [17]. (a) The dependence of the intensity at D2 on the cavity detuning. (b) The amounts of amplitude-sum squeezing after EOT from samples with different reflectivities, (i) without feedback control and (ii) with feedback control.

Our experiment has presented clear evidences that CV entanglement can survive going through a photon-plasmon-photon conversion process. With the help of quantum plasmons which inherit the quantum correlations of two beams, the basic CV-entangled characteristics can be well preserved. Because the uses of CVs offer the possibility of deterministically generating entangled states that can be measured with high efficiency, which is crucial for the efficient implementation of a number of quantum information protocols [14], the deterministic CV-entangled plasmons should be an attractive alternative to discrete-variable entangled systems in integrated quantum circuits. The exact physics behind the excitation of CV-entangled plasmons needs to be better clarified in future experiments. For instance, how can one depict the inheritance of the noise correlation in quantum plasmons? One potential way is to develop novel tools that can directly measure the quantum correlations of the plasmons. The sensitivity to losses for the CV-entangled states has always been the major worrying and limiting factor for its potential applications, and metals are known to have high losses. We have theoretically proposed and experimentally demonstrated a new feedback method to recover the reduced quantum correlations due to the reflected light from the sample. With further optimization, such coherent feedback method can be more effectively used to preserve the quantum correlations in plasmonic circuits, especially for the cases when it is impossible to design functionalized metal structures with low reflectivity.

In conclusion, quantum plasmons excited by CV-entangled states have been experimentally realized. The reflected light from the metal-hole array resonating in a weak feedback cavity is reused to further enhance the transmitted quantum entanglement. A stable feedback-control scheme is very important in the system, which would be easier to realize for the on-chip applications since the disturbance from the environment can be basically eliminated in integrated circuits. The coherent feedback control method demonstrated in this work provides an improved method for the integrated quantum plasmonic devices and on-chip measurements beyond the SNL [27].

This work was supported by National Basic Research Program of China (Nos. 2012CB921804 and 2011CBA00205), the National Natural Science Foundation of China (Nos. 11321063, 61205115, 11274162, and 61222503) and the Natural Science Research Program of Higher Education Institution of Anhui Province (No. KJ2012Z023). The authors thank Institute of Opto-Electronics of Shanxi University for providing the entanglement source. The authors also thank Lijian Zhang and Tao Li for useful discussions.